\documentclass[12pt,twoside]{article}
\usepackage{amsmath}

\newcommand{\bra}[1]{\left\langle #1 \right|}
\newcommand{\ket}[1]{\left|#1\right\rangle}
\newcommand{\vx}{\vec{x}}
\newcommand{\vX}{\vec{X}}
\newcommand{\vxi}{\vec{\xi}}
\newcommand{\vy}{\vec{y}}
\newcommand{\vz}{\vec{z}}
\newcommand{\vk}{\vec{k}}
\newcommand{\vp}{\vec{p}}
\newcommand{\vq}{\vec{q}}
\newcommand{\vw}{\vec{w}}
\newcommand{\M}{\mathcal{M}}
\newcommand{\W}{\mathcal{W}}
\newcommand{\V}{\mathcal{V}}
\begin{document}
\title{Photon emission rate from atomic systems in the CSL model}
\author{
Stephen L. Adler \\
\small Institute for Advanced Study\\
\small Einstein Drive\\
\vspace{.2in}
\small Princeton, NJ 08540\\
Fethi M. Ramazano\v glu\\
\small Department of Physics\\
\small Princeton University\\
\small Princeton, NJ 08544\\[-0.25in]}
\date{}
\maketitle \pagestyle{myheadings} \markboth{Adler and
Ramazanoglu}{photon emission rate from atomic systems in CSL
model} \thispagestyle{empty}
\begin{abstract}
\noindent  We calculate the photon emission rate from a general
atomic system in the mass-proportional CSL model. For an isolated
charged particle emitting kilovolt gamma rays, our results agree
with those obtained by Fu. For a neutral atomic system, photon
emission is strongly suppressed for photon wavelengths much larger
than the atomic radius.  However, for kilovolt gamma rays, Fu's
result is modified by a structure factor that is of order unity,
giving no rate suppression.  Our calculation is readily
generalized to the case of non-white noise, noise couplings that
are not mass-proportional, and general (non-Gaussian) spatial
correlation functions, and corresponding results are given.  We
briefly discuss the implications of our calculation for upper
bounds on the CSL model parameters.
\end{abstract}

\section{Introduction}
Stochastic modifications of the Schr\"odinger equation, such as
the continuous spontaneous localization (CSL) model, solve the
measurement problem in quantum theory by giving an objective
account of state vector reduction [1].  To assess the viability of
these models, it is necessary to estimate lower and upper bounds
on the stochastic model parameters, as surveyed in a recent paper
of Adler [2].  An important upper bound on the stochastic rate
parameter comes from a calculation by Fu [3] of the rate of
noise-induced gamma radiation from free electrons, which he
compares with the observed bound on 11 kilovolt gamma radiation
from germanium.   Adler suggested in [2] that in a neutral atomic
system, radiation from protons, in the case of mass-proportional
noise couplings, will largely cancel the radiation from electrons.
Our aim in this paper is to check this assertion by a detailed
calculation of stochastic noise-induced radiation in a general
atomic system.  We find that the asserted cancellation is present
only for very long wavelength photons, whereas for the 11 kilovolt
gamma rays figuring in Fu's bound, the radiation from protons
somewhat enhances, rather than reducing, that from electrons. This
result can be simply understood as the effect of inclusion of the
space coordinate-dependent phase factor for the radiated wave.

Thus for white noise, the upper bound on the CSL rate parameter is
six orders of magnitude lower than estimated in [2], and hence is
three orders of magnitude smaller that the lower bounds estimated
in [2] from processes of latent image formation, assuming that
latent image formation (and not subsequent development)
corresponds to state vector reduction.   Hence if the assumptions
on which these lower bounds are based are correct, the white noise
CSL model is disfavored. White noise is of course an idealization,
and our calculation can be readily extended to the case of
non-white noise.   For non-white noise with a spectral cutoff
below 11 kilovolts, there is no 11 kilovolt gamma radiation, and
so in this case the germanium experiment does not set a bound on
the CSL model rate parameter, and there is no conflict with the
lower bounds estimated in [2].

This paper is organized as follows.   In Sec. 2 we outline the
basic strategy of the calculation, which is to replace the real
noise of the CSL model by an imaginary noise, that can be
represented by a perturbation term in the Hamiltonian. We write
down the general form of the Hamiltonian, and give the noise
structure in the white-noise and non-white noise cases.

In Sec. 3 we use standard atomic physics methods [4] to derive a
master formula for the noise-induced photon radiation rate, in
both the white noise and the non-white noise cases.  In Sec. 4 we
evaluate this formula for a single free electron, recovering the
result of Fu [3] when his approximations are made. In Sec. 5 we
evaluate the master formula for a hydrogenic atom, and in Sec. 6
for a general atomic system.  In Sec. 7 we state the
generalization of our results to a noise perturbation with general
(not necessarily mass proportional) couplings to the particles,
and with general spatial and time correlation functions.  We
conclude with a brief discussion of the implications of our
calculation for CSL model phenomenology.

\section{General strategy, Hamiltonian, and noise structure}
In the CSL model, the stochastic Schr\"odinger equation obeyed by
the wave function $\psi$ takes the form $d\psi=-(i/\hbar)H\psi dt
+ {\cal N} \psi+...$, with $H$ the usual Hamiltonian, with the
noise term ${\cal N}$ real valued, and with the ellipsis ...
representing additional nonlinear terms needed to preserve state
vector normalization.   A real valued choice for the noise term
corresponds to an imaginary addition to the Hamiltonian, and is
necessary to obtain a  model that describes state vector
reduction.  An alternative stochastic Schr\"odinger equation can
be written with an imaginary noise term, which does not require
additional nonlinear terms in the Schr\"odinger equation for norm
preservation.   This Schr\"odinger evolution does not lead to
state vector reduction, but for the case of white noise, it is a
well known result that the noise average of the density matrix
obeys the same evolution equation in the real and imaginary noise
cases. Since the mean rate for noise induced transitions can be
calculated from the noise averaged density matrix, this implies
that one can use the imaginary noise equation to calculate the
mean rate for such transitions. Hence, to leading order, one can
represent the noise perturbation as a self-adjoint perturbation on
the Hamiltonian $H$, and use standard second order perturbation
theory to evaluate its effects.

The usual justification for the use of imaginary noise is based on
a calculation of the density matrix evolution in the real and
imaginary noise cases using the It\^o calculus, which as already
noted, assumes white noise. Adler and Bassi [5] have recently
shown, however, that in the case of non-white Gaussian noise, the
noise averaged density matrix evolutions are still the same for
the real and imaginary noise cases, through second order in the
noise term. Hence, in the second order perturbation calculations
of this paper, we can use an imaginary noise term to calculate the
effects of non-white noise as well as white noise.

We will thus be considering a Hamiltonian of the form
\begin{equation}\label{eq:basicham}
H = H_0 + H_{em} + H_n~~~,
\end{equation}
with $H_0$ the atomic system Hamiltonian, $H_{em}$ the
electromagnetic perturbation describing photon emission, and $H_n$
the perturbation describing the noise. For a system of $N$
particles of charges $e_j$ and masses $m_j$, the electromagnetic
perturbation is
\begin{equation}\label{eq:empert1}
H_{em} = \sum_{j=1}^N \frac{ie_j\hbar}{m_jc} \vec{A}(\vx_j) \cdot
\vec{\nabla}_{x_j} + O(\vec{A}^{\,2})~~~,
\end{equation}
with the electromagnetic potential, for field quantization in a
cubical box of size $L$, given by
\begin{equation}\label{eq:quantA}
\vec{A}(\vec{x}) = \sum_{\vp} \sqrt{\frac{2\pi \hbar c^2}{
\omega_p L^3}} \left[ a_p \vec{\epsilon}_p e^{i(\vp \cdot \vx -
\omega_p t)} + a_p^{\dag} \vec{\epsilon}_p e^{-i(\vp \cdot \vx -
\omega_p t)} \right]~~~,
\end{equation}
where $\omega_p=pc$, and where the numerical value of a unit
unrationalized
charge $e$ is $e^2/(\hbar c)\simeq 1/137.04$.   Since we are
only interested in the matrix element for emitting a single photon
of wave number $\vp$, we pull this term out from
Eq.~\ref{eq:quantA} and, separating off the time dependence, write
the electromagnetic perturbation as
\begin{equation}\label{eq:empert2}
\begin{split}
H_{em}=&e^{i\omega_p t} \W^p(\{\vx\})~~~,\\
\W^p(\{\vx\}) & = a_p^{\dag} \sqrt{\frac{2\pi \hbar c^2}{\omega_p
L^3}} \, \sum_j\frac{i e_j \hbar}{m_j c} e^{-i \vec{p} \cdot
\vx_j} \vec{\epsilon}_p \cdot \vec{\nabla}_j~~~,
\end{split}
\end{equation}
where $\vec{\nabla}_j$ is an abbreviation for
$\vec{\nabla}_{x_j}$.

In the CSL model with mass-proportional couplings, the noise
perturbation can be written as
\begin{equation}\label{eq:noisepert}
\begin{split}
H_{n}=&\int d^3z \frac{dW_t(\vz)}{dt}\V(\vz,\{x\})~~~,\\
\V(\vz,\{\vx\})=&-\frac{\hbar}{m_N}\sum_j m_j g(\vz-\vx_j)~~~.
\end{split}
\end{equation}
Here $g(\vx)$ is a spatial correlation function, conventionally
taken as the Gaussian
\begin{equation}\label{eq:correlfn}
g(\vx) = \left( \frac{\alpha}{2\pi} \right)^{3/2} e^{-\alpha
\vx^2/2} = \left( \sqrt{2\pi} r_c \right)^{-3} e^{-\vx^2/2
r_c^2}~~~,
\end{equation}
with $\alpha^{-{1\over 2}}=r_c$, and with $r_c$ conventionally
taken as $10^{-5}$ cm.   In the case of white noise, $dW_t$ is an
It\^o calculus differential that obeys
\begin{equation}\label{eq:whitenoise}
dW_t(\vx)dW_t(\vy) = \gamma dt \delta^3(\vx-\vy)~~~,
\end{equation}
with $\gamma$ the noise strength parameter. The corresponding
formula for the case of non-white noise is
\begin{equation}\label{eq:nonwhitenoise}
E\left[ \frac{dW_t(\vx)}{dt} \frac{dW_{t'}(\vy)}{dt'} \right] =
\frac{1}{2 \pi} \int_{-\infty}^{\infty} d\omega \, \gamma(\omega)
e^{-i \omega(t-t')} \delta^3(\vx-\vy)~~~,
\end{equation}
with $E[...]$ denoting the expectation or average over the noise.
When $\gamma(\omega)$ is a constant $\gamma$,
Eq.~\ref{eq:nonwhitenoise} reduces, on integration over
$t^{\prime}$, to Eq.~\ref{eq:whitenoise}.

\section{Master equation for the radiation rate}

According to Eqs.~\ref{eq:empert2} and \ref{eq:noisepert}, the
total perturbation on the atomic Hamiltonian $H_0$ is
\begin{equation}\label{eq:pertpot}
V(t) = \int d^3z \frac{dW_t(\vec{z})}{dt} \V(\vec{z},\{ \vx \}) +
e^{i \omega_p t} \W^p(\{ \vx\})~~~.
\end{equation}
Expanding the transition amplitude in a perturbation series
following the methods of [4], we get
\begin{equation}\label{eq:pertseries}
\begin{split}
\bra{f}U_I(t,0)\ket{i} & =1 + \mathcal{T}_{fi}^{(1)} +
\mathcal{T}_{fi}^{(2)} + ...~~~,\\
\mathcal{T}_{fi}^{(2)} & = -\frac{1}{\hbar^2} \int_0^t ds \int_0^s
du
\sum_k \bra{f}V_I(s)\ket{k} \bra{k} V_I(u)\ket{i}\\
& = -\frac{i}{2 \pi \hbar^2} \int_0^t ds \int_0^t du
\int_{-\infty}^{\infty} dE \, e^{\frac{i}{\hbar}(E_f-E)s}
e^{\frac{i}{\hbar}(E-E_i)u} \sum_k \frac{V_{fk}(s)V_{ki}(u)}{E_i+
i \eta-E_k}~~~,
\end{split}
\end{equation}
where in the first line of the formula for
$\mathcal{T}_{fi}^{(2)}$,  $V_I$ denotes the interaction picture
perturbation, and in the second line $V_{fk}$ and $V_{ki}$ denote
matrix elements of the Schr\"odinger picture perturbation. To
calculate the noise induced radiation, we are only interested in
the terms in Eq.~\ref{eq:pertseries} that are bilinear in the
electromagnetic and noise perturbations, so on substituting Eq.~
\ref{eq:pertpot} and dropping irrelevant terms, we get
\begin{equation}\label{eq:2nd_order_matrix_element}
\begin{split}
\mathcal{T}_{fi}^{(2)} & = \frac{-i}{2 \pi \hbar^2}
\int_{-\infty}^{\infty} dE \\
& \times\left( \int_0^t ds \thinspace e^{\frac{i}{\hbar}(E_f-E)s}
\int d^3z \frac{dW_s(\vz)}{ds} \int_0^t du \thinspace
e^{\frac{i}{\hbar}(E-E_i+\hbar
\omega_p)u} \sum_k \frac{\V_{fk}(\vz)\W^p_{ki}}{E+i \eta -E_k} \right. \\
+  & \left. \int_0^t  ds \thinspace e^{\frac{i}{\hbar}(E_f-E+\hbar
\omega_p)s} \int_0^t du \thinspace e^{\frac{i}{\hbar}(E-E_i)u}
\int d^3z \frac{dW_u(\vz)}{du} \sum_k \frac{\W^p_{fk} \V_{ki}
(\vz)}{E+i \eta -E_k}  \right)~~~.
\end{split}
\end{equation}
Taking the squared modulus of
Eq.~\ref{eq:2nd_order_matrix_element}, averaging over the noise,
and using the formulas for representations of the Dirac delta
function given in  [4], in the large time limit we obtain in the
white noise case,
\begin{equation}\label{eq:2nd_order_white_noise}
E[|\mathcal{T}_{fi}^{(2)}|^2] = \frac{\gamma t}{\hbar^2} \int d^3z
\left| \sum_k \frac{\V_{fk}(\vz)\W^p_{ki}}{E_i - \hbar \omega_p +i
\eta -E_k} + \frac{\W^p_{fk}\V_{ki}(\vz)}{E_f + \hbar \omega_p +i
\eta -E_k} \right|^2~~~,
\end{equation}
with the corresponding equation in the non-white noise case taking
the form
\begin{equation}\label{eq:2nd_order_nonwhite_noise}
E[|\mathcal{T}_{fi}^{(2)}|^2] = \frac{t}{\hbar^2} \gamma(
\omega_p+ \frac{E_f - E_i}{\hbar}) \int d^3z \left| \sum_k
\frac{\V_{fk}(\vz)\W^p_{ki}}{E_i - \hbar \omega_p +i \eta -E_k} +
\frac{\W^p_{fk} \V_{ki}(\vz)}{E_f + \hbar \omega_p +i \eta -E_k}
\right|^2~~~.
\end{equation}
Equations ~\ref{eq:2nd_order_white_noise} and
~\ref{eq:2nd_order_nonwhite_noise} are the master equations from
which we shall calculate the noise induced radiation rate, by
substituting the matrix elements of $\V$ and $\W^p$ appropriate to
the various cases of interest.

\section{Free electron: repeating Fu's calculation}
As a first application of Eq.~\ref{eq:2nd_order_white_noise}, and
a check, let us repeat the calculation of Fu [3] for the case of a
single free electron.   Assuming that the electron is initially at
rest, the initial, final, and intermediate state electron wave
functions are
\begin{equation}\label{eq:single_particle_wavefunctions}
\psi_i=\frac{1}{\sqrt{L^3}}~~, \qquad \psi_f=\frac{e^{i \vq \cdot
\vx}}{\sqrt{L^3}}~~, \qquad \psi_k=\frac{e^{i \vk \cdot
\vx}}{\sqrt{L^3}}~~~.
\end{equation}
{}From Eqs.~4 and 5, as specialized to a single particle of charge
$e$ (with $e^2/(\hbar c) \simeq 1/137$) and mass $m$, the needed
matrix elements are
\begin{equation}\label{eq:free_electron_W}
\begin{split}
\W^p_{ki} & = 0~~~,\\
\W^p_{fk} & = -\sqrt{\frac{2\pi \hbar c}{p L^3}} \, \frac{e
\hbar}{m c} \vec{\epsilon}_p \cdot \vq \, \delta_{\vk
-\vec{p}-\vq}~~~,
\end{split}
\end{equation}
and
\begin{equation}\label{eq:free_electron_V}
\begin{split}
\V_{ki}(z) & = - \frac{\hbar m}{m_N L^3} \, e^{-i \vk \cdot \vz
-\frac{1}{2} \vk^2 r_c^2}~~~,\\
\V_{fk}(z) & = - \frac{\hbar m}{m_N L^3} \, e^{i(\vk-\vq) \cdot
\vz -\frac{1}{2}(\vk-\vq)^2r_c^2}~~~.
\end{split}
\end{equation}

Substituting these into Eq.~\ref{eq:2nd_order_white_noise}, we get
for the noise averaged squared matrix element
\begin{equation}\label{eq:single_particle_T^2}
E[|\mathcal{T}_{fi}^{(2)}|^2] = \frac{\gamma t}{\hbar^2} \int d^3z
\left| \frac{\hbar}{m_N L^3} \sqrt{\frac{2\pi \hbar c}{p L^3}} \,
\frac{e \hbar}{c} \, \vec{\epsilon}_p \cdot \vq \,
\frac{e^{-i(\vec{p}+\vq) \cdot \vz -\frac{1}{2}
(\vec{p}+\vq)^2r_c^2}}{-\frac{\hbar^2}{2m} (p^2+2 \vec{p} \cdot
\vq) +\hbar c p +i \eta} \right|^2~~~.
\end{equation}

Fu notes that when the photon momentum $p$ is much larger than the
inverse correlation length $1/r_c$, the Gaussian factor in
Eq.~\ref{eq:single_particle_T^2} forces the electron and photon to
emerge nearly back to back, that is, $\vq\simeq -\vp$. As a result
\begin{equation}\label{eq:approx}
\frac{\hbar c p}{-\frac{\hbar^2}{2m}(p^2+2\vp \cdot \vq)}\simeq
\frac{2 m c^2}{\hbar pc}~~~,
\end{equation}
which is of order 100 for $\hbar pc=$ 11 keV.  Thus one can to a
good approximation keep only the term $\hbar c p$ in the
denominator of Eq.~\ref{eq:single_particle_T^2}, which then
simplifies to
\begin{equation}\label{eq:single_particle_T^2_approx}
E[|\mathcal{T}_{fi}^{(2)}|^2] = \frac{\gamma t}{\hbar^2}
\left(\frac{\hbar}{m_N L^3}\right)^2  \frac{2\pi \hbar c}{p} \,
\frac{e^2 }{c^4 p^2} \left( \vec{\epsilon}_p \cdot \vq \right)^2
e^{-(\vec{p}+\vq)^2 r_c^2}~~~.
\end{equation}
Integrating over phase space for the electron and photon, summing
over photon polarizations, and dividing by the elapsed time,  we
get for the  radiated power per unit photon momentum space volume
and per unit time,
\begin{equation}\label{eq:electron_power1}
\frac{dP}{d^3p} = \left( \frac{L}{2\pi} \right)^6 \int d^3q
\sum_{\epsilon} E[|\mathcal{T}_{fi}^{(2)}|^2] \frac{1}{t}~~~.
\end{equation}
Carrying out the integrals and polarization sum, and replacing the
noise parameter $\gamma$ by a new parameter $\lambda$ defined by
$\gamma =8 \pi^{3/2} r_c^3 \lambda$, we get finally for the power
radiation rate
\begin{equation}\label{eq:electron_power2}
\frac{dP}{dp} = \frac{\hbar}{c^3} \, \frac{e^2 \lambda}{\pi r_c^2
m_N^2 p}~~~.
\end{equation}
This is in agreement with the result obtained by Fu [3], when our
unrationalized charge squared $e^2$ is replaced by $e^2/(4\pi)$,
corresponding to Fu's use of a rationalized charge convention.

\section{Hydrogenic atom}
We consider next a hydrogenic atom, with oppositely charged
particles of masses $m_1$ and $m_2$.  Equation 5 for $\V(\vz,
\{\vx\})$ now takes the form
\begin{align}\label{eq:perturbation_terms_hydrogen}
\V(\vz,\{\vx\})=&-\frac{\hbar}{m_N}\M(\vz,\{\vx\})~~~,\\
\M(\vz,\{\vx\})=&m_1g(\vz-\vx_1)+m_2g(\vz-\vx_2)~~~.
\end{align}
Introducing the center of mass coordinate $\vX$, total mass $M$,
relative coordinate $\vx$, and reduced mass $\mu$, by
\begin{align}\label{eq:CoM_coordinates_hydrogen}
\vec{X} =& \frac{m_1}{M}\vx_1 + \frac{m_2}{M}\vx_2, \quad \quad
\quad \vx = \vx_1 -\vx_2~~~,\\
M=&m_1+m_2, \quad \quad \mu=\frac{m_1m_2}{M}~~~,
\end{align}
 we can use the fact that the Bohr radius $a_0$ is much smaller
than $r_c$ to approximate  $\M(\vz,\{\vx\})$ as follows,
\begin{equation}\label{eq:M_hydrogen}
\begin{split}
\M(\vz,\{\vx\}) &=  m_1g(\vz-\vx_1) +  m_2 g(\vz-\vx_2)\\
& =M \, g(\vz-\vec{X}) + \frac{m_1m_2}{2M}
\, (\vx \cdot \vec{\nabla}_z)^2g(\vz-\vec{X})\\
& \cong M \, g(\vz-\vec{X})~~~,
\end{split}
\end{equation}
giving
\begin{align}\label{eq:hydrogen_V_W}
\W^p(\{\vx\}) &= a_p^{\dagger}\sqrt{\frac{2\pi \hbar c}{pL^3}} \,
\frac{ie\hbar}{c} \, \vec{\epsilon}_p \cdot \left( \frac{1}{m_1}
\, e^{-i\vec{p} \cdot \vx_1} \, \vec{\nabla}_1 -\frac{1}{m_2} \,
e^{-i\vec{p} \cdot \vx_2}
\, \vec{\nabla}_2 \right)~~~,\\
\V(\vz,\{\vx\}) &\cong - \frac{\hbar M}{m_N} \, g(\vz-\vec{X})~~~.
\end{align}
The initial, final, and itermediate state atomic wave functions
are now
\begin{equation}\label{eq:hydrogen_wavefunctions}
\psi_i=\frac{1}{\sqrt{L^3}}u_{\hat{i}}(\vx)~~, \qquad
\psi_f=\frac{e^{i \vq \cdot \vX}}{\sqrt{L^3}}u_{\hat{f}}(\vx)~~~,
\qquad \psi_k=\frac{e^{i \vk \cdot \vX}}{\sqrt{L^3}}u_{\hat{
k}}(\vx)~~~,
\end{equation}
where we use carets to denote the labels of hydrogenic internal
states. Defining
\begin{equation}\label{eq:O_operator}
\mathcal{O}(\vk) = \frac{i}{M} \left( e^{-i\frac{m_2}{M}\vec{p}
\cdot \vx} - e^{i \frac{m_1}{M}\vec{p} \cdot \vx} \right)
\vec{\epsilon}_p \cdot \vk + \left( \frac{1}{m_1}
e^{-i\frac{m_2}{M}\vec{p} \cdot \vx} +\frac{1}{m_2}
e^{i\frac{m_1}{M}\vec{p} \cdot \vx} \right) \vec{\epsilon}_p \cdot
\vec{\nabla}_x~~~,
\end{equation}
we find that the matrix elements entering the master formula are
\begin{align}\label{eq:hydrogen_W_no_approximation}
\W^p_{ki} & = \sqrt{\frac{2\pi \hbar c}{pL^3}} \, \frac{ie\hbar}{c}
\bra{\hat{k}}\mathcal{O}(\vec{0}) \ket{\hat{i}}
\, \delta_{\vk+\vec{p}}\\
\W^p_{fk} & = \sqrt{\frac{2\pi \hbar c}{pL^3}} \,
\frac{ie\hbar}{c} \bra{\hat{f}}\mathcal{O}(\vk) \ket{\hat{k}} \,
\delta_{\vk - \vec{p}-\vq}~~~,
\end{align}
and
\begin{align}\label{eq:hydrogen_V}
\V_{ki} &=  -\frac{\hbar M}{m_NL^3} \, e^{-i \vk \cdot \vz -
\frac{1}{2} \vk^2 r_c^2} \, \delta_{\hat{k} \hat{i}}\\
\V_{fk} &=  -\frac{\hbar M}{m_NL^3} \, e^{i (\vk-\vq) \cdot \vz -
\frac{1}{2} (\vk-\vq)^2 r_c^2} \,  \delta_{\hat{f} \hat{k}}~~~.
\end{align}
Then without any further approximation we find
\begin{multline}\label{eq:hydrogen_T^2_no_approximation}
E[|\mathcal{T}_{fi}^{(2)}|^2] = \frac{\gamma t}{\hbar^2}
\left(\frac{\hbar M}{m_N L^3}\right)^2  \frac{2\pi \hbar c}{p} \,
\frac{e^2 \hbar^2}{c^2} e^{-(\vec{p}+\vq)^2 r_c^2}\\
\times \left| - \frac{\bra{\hat{f}} \mathcal{O}(\vec{0})
\ket{\hat{i}}}{E_{fi}+\frac{\hbar^2p^2}{2M} + \hbar c p -i\eta} +
 \frac{\bra{\hat{f}} \mathcal{O}(\vec{p}+\vq)
\ket{\hat{i}}}{E_{fi}+\frac{\hbar^2q^2}{2M} + \hbar c p
-\frac{\hbar^2 (\vec{p}+\vq)^2}{2M} +i \eta} \right|^2~~~,
\end{multline}
with $E_{fi}\equiv E_{\hat{f}}-E_{\hat{i}}$ the internal energy
difference between the final and initial atomic states.   The
radiated power, per unit photon momentum space volume and per unit
time, now requires a sum over the final internal atomic state
$\hat{f}$, and is given by
\begin{equation}\label{eq:hydrogen_radiation_power}
\frac{dP}{d^3p} = \left( \frac{L}{2\pi} \right)^6 \int d^3q
\sum_{\hat{f},\, \epsilon} E[|\mathcal{T}_{fi}^{(2)}|^2]
\frac{1}{t}~~~.
\end{equation}

Note that when $\vp+\vq=0$, the two terms in Eq. 37 cancel. Since
the Gaussian $e^{-(\vp+\vq)^2 r_c^2}$ constrains $|\vp+\vq|$ to be
not much larger than $1/r_c$, we can make this cancellation
explicit by expanding in the small parameter
\begin{equation}\label{eq:hydrogen_series_expansion_term}
\frac{\hbar^2 \vec{p} \cdot (\vec{p}+\vq)}{M \left(\hbar c p +
\frac{\hbar^2 p^2}{2M} + E_{fi} \right)} \equiv \frac{\hbar^2
\vec{p} \cdot (\vp+\vq)}{M D_0}~~~,
\end{equation}
which keeping the leading two terms, and writing $\vp \cdot \vx
=pz$,  gives
\begin{equation}\label{eq:hydrogen_radiation_power_series_expanded}
\begin{split}
\frac{dP}{dp} =& p \, \frac{\hbar^3}{c}
\left(\frac{M}{m_N}\right)^2 \frac{e^2 \lambda}{\pi r_c^2}
\sum_{\hat{f}} \left\{ \, \frac{1}{M^2D_0^2} \left| \bra{\hat{f}}
e^{-i\frac{m_2}{M} pz} - e^{i \frac{m_1}{M} pz}
\ket{\hat{i}} \right|^2 \right. \\
+& \left. \frac{p^2\hbar^4}{M^2D_0^4} \left| \bra{\hat{f}} \left(
\frac{1}{m_1} e^{-i\frac{m_2}{M} pz} +\frac{1}{m_2}
e^{i\frac{m_1}{M} pz} \right) \frac{\partial}{\partial x}
\ket{\hat{i}} \right|^2 \right\}~~~.
\end{split}
\end{equation}
For small $p$, this expression can be further simplified to
\begin{equation}\label{eq:hydrogen_radiation_small_p}
\begin{split}
\frac{dP}{dp} & = p^3 \, \frac{\hbar^3}{c}
\left(\frac{M}{m_N}\right)^2 \frac{e^2 \lambda}{\pi r_c^2}
\sum_{\hat{f}} \left\{ \, \frac{1}{M^2E_{fi}^2} \left|
\bra{\hat{f}} z \ket{\hat{i}} \right|^2 + \frac{\hbar^4}{M^2
E_{fi}^4\mu^2} \left| \bra{\hat{f}} \frac{\partial}{\partial x}
\ket{\hat{i}} \right|^2
\right\}\\
& =2\, p^3 \,\frac{\hbar^3}{c} \, \frac{1}{m_N^2} \, \frac{e^2
\lambda}{\pi r_c^2} \sum_{\hat{f}} \frac{\left| \bra{\hat{f}} z
\ket{\hat{i}} \right|^2}{E_{fi}^2}
\end{split}
\end{equation}
where we have used the dipole approximation formula
\begin{equation}\label{eq:p_expectation_to_x_expectation}
\left| \bra{\hat{f}} \frac{\partial}{\partial x} \ket{\hat{i}}
\right| =  \frac{\mu E_{fi}}{\hbar^2} \left| \bra{\hat{f}} x
\ket{\hat{i}} \right|~~~,
\end{equation}
which shows that the two terms in
Eq.~\ref{eq:hydrogen_radiation_small_p} make equal contributions.
The sum in Eq.~\ref{eq:hydrogen_radiation_small_p} has been
evaluated in closed form by Dalgarno and Kingston [6], with the
result
\begin{equation}\label{eq:sum}
 \sum_{\hat{f}} \frac{\left| \bra{\hat{f}} z \ket{\hat{i}}
\right|^2}{E_{fi}^2}=\frac{43}{8} \frac{\mu^2 a_0^6} {\hbar^4}~~~,
\end{equation}
giving an explicit expression for the small $p$ radiation rate.

However, for 11 kilovolt photons, the small $p$ approximation does
not apply, and instead we can simplify the formulas by making the
approximation $D_0 \approx \hbar c p$, as was done by Fu in his
calculation.  The radiation rate then becomes
\begin{multline}\label{eq:hydrogen_radiation_high_p}
\frac{dP}{dp} = \frac{1}{p} \, \frac{\hbar}{c^3}
\left(\frac{M}{m_N}\right)^2 \frac{e^2 \lambda}{\pi r_c^2}
\sum_{\hat{f}} \left\{ \, \frac{1}{M^2} \left| \bra{\hat{f}}
e^{-i\frac{m_2}{M} pz} - e^{i \frac{m_1}{M} pz}
\ket{\hat{i}} \right|^2 \right. \\
+ \left. \frac{\hbar^2}{M^2 c^2} \left| \bra{\hat{f}} \left(
\frac{1}{m_1} e^{-i\frac{m_2}{M} pz} +\frac{1}{m_2}
e^{i\frac{m_1}{M} pz} \right) \frac{\partial}{\partial x}
\ket{\hat{i}} \right|^2 \right\}~~~,
\end{multline}
which, using the completeness of the hydrogen spectrum, can be
simplified to
\begin{multline}\label{eq:hydrogen_radiation_high_p_simplified}
\frac{dP}{dp} = \frac{1}{p} \, \frac{\hbar}{c^3}
\left(\frac{M}{m_N}\right)^2 \frac{e^2 \lambda}{\pi r_c^2}
\sum_{\hat{f}} \left\{ \, \frac{1}{M^2}  \bra{\hat{i}} 2-2
\cos pz \ket{\hat{i}}  \right. \\
+ \left. \frac{\hbar^2}{M^2 c^2}  \bra{\hat{i}}
\frac{\overleftarrow{\partial}}{\partial x} \left( \frac{1}{m_1^2}
+ \frac{1}{m_2^2} +\frac{2}{m_1 m_2} \cos pz\right)
\frac{\partial}{\partial x} \ket{\hat{i}} \right\}~~~.
\end{multline}

The ratio of the second term to the first can be shown to be of
order $(e^2/\hbar c)^2$, so the second term can be neglected.
Evaluating the first term using the hydrogen atom ground state
wave function, we find the final result for high $p$ to be
\begin{equation}\label{eq:hydrogen_radiation_high_p_final}
\frac{dP}{dp} = 2 \left[ 1-\frac{1}{\left[ 1+\left(\frac{p a_0}{2}
\right)^2 \right]^2} \right] \frac{1}{p} \,\frac{\hbar}{c^3} \,
\frac{1}{m_N^2} \, \frac{e^2 \lambda}{\pi r_c^2}~~~.
\end{equation}
For small $p$ this expression is suppressed with respect to the
rate calculated by Fu, but for large $p$ it approaches twice Fu's
rate, because when the photon wave length is much smaller than the
atomic radius, the electron and proton radiation rates add
incoherently.   For 11 kilovolt gamma radiation
from hydrogen, the rate given by
Eq.~\ref{eq:hydrogen_radiation_high_p_final} is about 1.8 times
the rate for a free electron. The structure of the first term in
Eqs.~\ref{eq:hydrogen_radiation_high_p}  and
~\ref{eq:hydrogen_radiation_high_p_simplified} can be readily
understood in terms of the phase factor that appears in the
formula for the radiation rate of a distributed charge system, as
in Eqs. (13-33) and (13-37) of the text of Panofsky and Phillips
[7].

\section{Many-body system}

We turn next to a general $n$ particle atomic system, for which
the electromagnetic and noise perturbations are given by Eqs. 4
and 5, with the sum over $j$ extending from 1 to $n$.  In order to
take account of overall momentum conservation, we separate the
coordinates of the particles into a center of mass coordinate
$\vX$ and internal coordinates $\vxi_i~,i=1,...,n-1$, by writing

\begin{equation}\label{eq:many_body_CoM_coordinates}
\begin{split}
M&=\sum_{j=1}^n m_j~~~,\\
\vx_i &= \vxi_i + \vX \quad \quad i=1,...,n-1~~~,\\
\vx_n &= \vX -\frac{1}{m_n} \sum_{j=1}^{n-1} m_j \vxi_j~~~,\\
\vxi_i &= \vx_i - \sum_{j=1}^n \frac{m_j \vx_j}{M}~~~,\\
\vX &= \sum_{j=1}^n \frac{m_j \vx_j}{M}~~~.
\end{split}
\end{equation}
In the following equations, $\vec{\nabla}_j$ denotes the partial
derivative with respect to the original coordinate $\vx_j$, not
the derivative with respect to  the internal coordinate $\vxi_j$.
Straightforward calculations show that the commutator of this
partial derivative with an internal coordinate is given by
\begin{equation}\label{eq:commutator}
[\vec{a}\cdot \vec{\nabla}_i,\vec{b}\cdot \vxi_j]=\vec{a}\cdot
\vec{b} (\delta_{ij}-\frac{m_i}{M})~~~,
\end{equation}
and also that the Jacobian $J$ of the transformation of Eq.
\ref{eq:many_body_CoM_coordinates} is given by
\begin{equation}\label{eq:jacobian}
J=\frac{\partial(\vx_1...\vx_n)}{\partial(\vX\vxi_1...\vxi_{n-1})}
=(-1)^{n-1}\left(1+\frac{\sum_{j=1}^{n-1} m_j}{m_n}\right)^3~~~.
\end{equation}
Moreover, the kinetic term of the unperturbed hamiltonian is
separated by the transformation of
Eq.~\ref{eq:many_body_CoM_coordinates} into a center of mass part
and an internal part,
\begin{equation}\label{eq:many-body_kinetic_energy_hamiltonian}
\sum_{i=1}^n  \frac{\vec{\nabla}_i^2}{2m_i} = \frac{\vec
\nabla_{X}^2}{2M} + \sum_{i=1}^{n-1} \frac{\vec
\nabla_{\xi_i}^2}{2m_i}  - \frac{1}{2M} \left( \sum_{i=1}^{n-1}
\vec \nabla_{\xi_i} \right)^2~~~,
\end{equation}
so that we know that wave functions are of the factorized form
\begin{equation}\label{eq:many-body_wavefunctions}
\psi_i=\frac{1}{\sqrt{L^3}} u_{\hat{i}}(\{ \vxi ~\}) \qquad
\psi_f=\frac{e^{i \vq \cdot \vX}}{\sqrt{L^3}} u_{\hat{f}}(\{ \vxi
~\}) \qquad \psi_k=\frac{e^{i \vk \cdot \vX}}{\sqrt{L^3}}
u_{\hat{k}}(\{ \vxi ~\})~~~.
\end{equation}

Using the center of mass transformation and the factorized wave
functions, the noise and radiation matrix elements needed for the
master formula of Eq. 12 are calculated to be
\begin{align}\label{eq:many-body_W_no_approximation}
\W^p_{ki} & = \sqrt{\frac{2\pi \hbar c}{pL^3}} \, \frac{i\hbar}{c}
\bra{\hat{k}} \sum_j e^{-i \vec{p} \,\cdot \vec{\xi}_j} e_j
\frac{\vec{\epsilon}_p}{m_j} \cdot \vec{\nabla}_j \ket{\hat{i}}
\, \delta_{\vk+\vec{p}}\\
\W^p_{fk} & = \sqrt{\frac{2\pi \hbar c}{pL^3}} \, \frac{i\hbar}{c}
\bra{\hat{f}} \sum_j e^{-i \vec{p}\, \cdot \vec{\xi}_j} e_j \left(
i \frac{\vec{\epsilon}_p \cdot \vk}{M} +
\frac{\vec{\epsilon}_p}{m_j} \cdot \vec{\nabla}_j \right)
\ket{\hat{k}} \, \delta_{\vk - \vec{p}-\vq}
\end{align}
and
\begin{align}\label{eq:many-body_V}
\V_{ki}(\vz) &=  -\frac{\hbar}{m_NL^3} \, e^{-i \vk \cdot \vz -
\frac{1}{2} \vk^2 r_c^2} \, \bra{\hat{k}} \sum_j e^{i \vk \cdot
\vxi_j} m_j \ket{\hat{i}}\\
\V_{fk}(\vz) &=  -\frac{\hbar}{m_NL^3} \, e^{i (\vk-\vq) \cdot \vz
- \frac{1}{2} (\vk-\vq)^2 r_c^2} \,  \bra{\hat{f}} \sum_j e^{-i
(\vk -\vq) \cdot \vxi_j} m_j \ket{\hat{k}}~~~.
\end{align}

We now simplify Eq. 12 by making the approximation that the photon
energy $\hbar \omega_p$ is much larger than both the internal
energy differences and the center of mass recoil energy, that is,
that $\hbar \omega_p$  is much larger than $E_i-E_k$ and
$E_f-E_k$. With this approximation (which is analogous to the
approximation made by Fu and also made in Eqs. 42-44 of our
hydrogen atom calculation), Eq.~12 simplifies to
\begin{equation}\label{eq:2nd_order_simplified}
E[|\mathcal{T}_{fi}^{(2)}|^2] = \frac{\gamma t}{\hbar^2} \int d^3z
\left| \sum_k \frac{ \V_{fk}(\vz)\W^p_{ki}-\W^p_{fk}\V_{ki}(\vz) }
{\hbar \omega_p}  \right|^2~~~.
\end{equation}
Substituting Eqs.~50-53 into Eq.~\ref{eq:2nd_order_simplified},
summing over the final state $f$ by the analog of Eq.~36, and
using completeness twice together with algebraic simplification
using Eq.~\ref{eq:commutator}, we get for the power radiated
\begin{equation}\label{eq:many-body_radiation_power}
\begin{split}
\frac{dP}{dp} &= \frac{2 \gamma}{(2\pi)^4} \,
\frac{\hbar}{m_N^2c^3} \, \frac{1}{p} \, \int \frac{d\Omega_
{\hat{p}}}{4\pi} \int d^3w \,e^{-\vec{w}^2 r_c^2}
[\vec{w}^2-(\vec{w} \cdot \hat{p})^2]
\bra{\hat{i}} |\mathcal{N}|^2 \ket{\hat{i}} \\
\mathcal{N} &= \sum_j e^{-i (\vec{p}-\vec{w}) \cdot \vec{\xi}_j} e_j
\end{split}
\end{equation}
Note that the internal integration to be used in evaluating the
matrix element in this formula includes the Jacobian $J$ of
Eq.~\ref{eq:jacobian}, and so is
\begin{equation}\label{eq:internalint}
|J|\prod_{j=1}^{n-1}\int d^3 \xi_j~~~.
\end{equation}

To check that Eq.~\ref{eq:many-body_radiation_power} reproduces
the result of the first term of Eq. 43 for the hydrogen atom, we
note first that for a two particle system one has
$\vx_1=\vX+\vxi_1\,, \vx_2=\vX+\vxi_2$, and so
$\vx=\vx_1-\vx_2=\vxi_1-\vxi_2$, which by
Eq.~\ref{eq:many_body_CoM_coordinates} reduces to $\vx=
\vxi_1(1+m_1/m_2)$.  Hence $|J|d^3 \xi_1 = (1+m_1/m_2)^3 d^3 \xi_1
=d^3 x$, so the internal integration involves the conventional
internal coordinate used for the hydrogen atom.  The expansion in
the small parameter of Eq.~37 is equivalent, in the many-body
context, to setting $\vw=0$ in $\mathcal{N}$ in
Eq.~\ref{eq:many-body_radiation_power}, an approximation that
permits the integration over $\vw$ to be easily done, yielding our
previous formula for the hydrogen atom radiated power.

One can also apply Eq.~\ref{eq:many-body_radiation_power} to the
case of a crystal lattice.  Again making the approximation of
neglecting $\vw$ in $\mathcal{N}$, that is, taking $r_c$ to be
large, we define
\begin{equation}\label{eq:single_cell_in_crystal}
f \equiv \sum_{cell} e^{-i \vec{p}\, \cdot \vec{\xi}_i} e_i~~~.
\end{equation}
We then find that the matrix element appearing in
Eq.~\ref{eq:many-body_radiation_power} takes the form (with
$\langle...\rangle$ denoting an expectation in the initial state
$\ket{\hat{i}} $, and with $\vec{L}_i$ a lattice displacement),
\begin{equation}\label{eq:many-body_N^2}
\begin{split}
\langle |\mathcal{N}|^2\rangle &= N_{cell} \left( \langle |f|^2
\rangle -|\langle f \rangle|^2\right) + \left| \sum_L e^{-i \vec{p}
\cdot \vec{L}_i} \right|^2 |\langle f \rangle|^2\\
& \cong N_{cell} \langle |f-\langle f \rangle |^2\rangle ~~~,
\end{split}
\end{equation}
since the second term on the first line of
Eq.~\ref{eq:many-body_N^2} grows more slowly than $N_{cell}$ for
generic values of $\vp$.  Hence as long as the variance of $f$
over a unit cell is nonzero, the radiated power scales as the size
of the crystal lattice (at least for lattice dimensions smaller
than $r_c$).

\section{Generalizations and discussion}
Several generalizations of the formulas given above can be easily
derived.  First of all, if the noise Hamliltonian of Eq. 5
involves general couplings $g_i$ that may differ from the masses
$m_i$, so that
\begin{equation}\label{eq:noisepertgen}
\begin{split}
H_{n}=&\int d^3z \frac{dW_t(\vz)}{dt}\V(\vz,\{x\})~~~,\\
\V(\vz,\{\vx\})=&-\frac{\hbar}{m_N}\sum_j g_j g(\vz-\vx_j)~~~,
\end{split}
\end{equation}
then in $\mathcal{N}$ in Eq.~\ref{eq:many-body_radiation_power}
one replaces $e_j$ by $e_j g_j/m_j$.  Secondly, our calculation,
in the non-white noise case, can be viewed as calculating the
radiation produced by a random gravitational potential
\begin{equation}\label{eq:randgrav}
V_{grav}(\vx,t)=\sum_i m_i \phi(\vx_i,t)~~~,
\end{equation}
with $\langle \phi\rangle_{AV}=0$ and with the correlation
function
\begin{align}\label{eq:corrfn}
\langle \phi(\vx,t)\phi(\vx^{\,\prime},t')\rangle_{AV}=&
\left(\frac{\hbar}{m_N}\right)^2
\frac{1}{2\pi}\int_{-\infty}^{\infty} d\omega \gamma(\omega)
e^{-i\omega(t-t')}G(\vx-\vx^{\,\prime})~~~,\\
G(\vx-\vx^{\,\prime})=&\int d^3z g(\vx-\vz) g(\vz-\vx^{\,\prime})
~~~.
\end{align}
Since for the Gaussian $g$ of Eq. 6 one has
\begin{align}\label{eq:fouriertrans}
\int d^3x e^{i\vk \cdot \vx}
g(\vx)=&e^{-\frac{1}{2}\vk^2r_c^2}~~~,\\
\int d^3x e^{i\vk \cdot \vx} G(\vx)=&\int d^3x e^{i\vk \cdot
\vx}\int d^3y g(\vx-\vy)g(\vy)=e^{-\vk^2r_c^2}~~~,
\end{align}
for a general $G(\vx)$ in Eq.~\ref{eq:corrfn} one simply replaces
$e^{-\vw^2 r_c^2}$ in the radiated power expressions by
\begin{equation}\label{eq:genfourtran}
G[\vw]=\int d^3x e^{i\vw \cdot \vx} G(\vx)~~~.
\end{equation}
Finally, for a more general non-white noise that does not have a
time-translation invariant correlation function, so that Eq. 8 is
replaced by
\begin{equation}\label{eq:gennonwhitenoise}
E\left[ \frac{dW_t(\vx)}{dt} \frac{dW_{t'}(\vy)}{dt'} \right] =
\Delta(t,t') \delta^3(\vx-\vy)~~~,
\end{equation}
the master formula in the non-white noise case is modified by
replacing
\begin{equation}\label{eq:orig}
t \gamma( \omega_p+ \frac{E_f - E_i}{\hbar})
\end{equation}
by
\begin{equation}\label{eq:gen}
\int_0^t ds \int_0^t du \Delta(s,u) e^{i(s-u)[ \omega_p+ \frac{E_f
- E_i}{\hbar}]}~~~.
\end{equation}
The most general case, in which the correlation function of Eq.~61
does not factorize into a temporal correlation times a spatial
correlation, can be obtained by combining results from Eqs. 61-68.

 To conclude, we consider the implications of our results for
CSL model phenomenology.  Since we have seen that for a hydrogenic
or a general atomic system emitting kilovolt gamma rays, charge
neutrality does not imply a corresponding cancellation in the
radiation rate, the estimates of Fu [3] must be taken as giving
the best upper bounds on the CSL parameter $\lambda$ (defined
following Eq.~20) for the white noise case.  Including [2] a
factor of $4\pi$ correction to Fu's evaluation of the electric
charge squared $e^2$, as well as [8] a factor of roughly 4 increase in
the experimental rate limit subsequent to the value used by Fu,
Fu's calculation implies
the bound $\lambda < 7 \times 10^{-11}\,{\rm s}^{-1}$, which is
$\sim 3 \times 10^{6}$ larger than the standard CSL model value of
$\lambda = 2.2 \times 10^{-17} {\rm s}^{-1}$. As we noted in Sec.
1, this upper bound is several orders of magnitude below the lower
bound on $\lambda$ set by postulating that latent image formation
(as opposed to image development) should correspond to state
vector reduction.  Although increasing $r_c$ to $10^{-4}$ cm
decreases the 11 kilovolt photon radiation rate, and so increases
the corresponding upper bound on $\lambda$, by two orders of
magnitude, as discussed in [2]  this increase in $r_c$ also
increases the latent image formation lower bound on $\lambda$ by
one to two orders of magnitude, and so does not eliminate the
potential discrepancy.

By contrast, in the non-white noise case there is not necessarily
a conflict, since the relevant radiation rate involves the noise
spectral coefficient $\gamma(\omega)$ at a frequency of at least
that of the emitted gamma ray, of order $10^{18}$$ {\rm s}^{-1}$
In fact, in their review [1], Bassi and Ghirardi suggest a cutoff
in the noise frequency spectrum  of order $c/r_c \sim 10^{15}$
${\rm s}^{-1}$, which would be more than sufficient. Even a much
lower frequency cutoff would suffice to explain reduction in
typical measurements with measurement times of order a nanosecond
or longer; for example, a cutoff of order $10^{11}$$ {\rm s}^{-1}$
would be more than adequate.   This would correspond to an energy
cutoff of order $10^{-4}$ eV, or a noise temperature of order $ 1$
degree K. So possibly even a non-white cosmic relic background
noise field, with suitable correlator structure, coupling as a
real-valued noise term ${\cal N}$ in the Schr\"odinger equation
for $d\psi$, could explain state vector reduction in measurement
situations, without coming close to violating the upper bound set
by Fu's calculation.

\section{Acknowledgments}
One of us (SLA) wishes to thank Philip Pearle and Angelo Bassi for
informative conversations.  The work of SLA was supported in part
by the Department of Energy under grant no DE-FG02-90ER40642, and
the manuscript was completed while he was at the Aspen Center for
Physics.  The other of us (FMR) was supported by the Piel Graduate
Fellowship of Princeton University.

\section{Added  Note}
The use of the term ``power'' and the symbol $P$ in Eqs. (20), (21),
(36), (38), (39), (42), (43), and (55) was inadvertent; we should
have used the term ``rate'' and the conventional symbol $\Gamma$.
These formulas all give the photon radiation rate, and do not
include the energy per photon factor $\hbar c p$ needed to convert
them to radiated power.  We wish to thank Angelo Bassi for pointing
this out to us.

\section{References}
\bigskip
[1] Bassi A and Ghirardi GC 2003 Dynamical reduction models {\it
Phys. Rep.} {\bf 379} 257; Pearle P 1999 Collapse models {\it Open
Systems and Measurements in Relativistic Quantum Mechanics
(Lecture Notes in Physics} vol 526) ed H-P Breuer and F
Petruccione (Berlin:Springer) \hfill\break
\medskip
[2] Adler S L  2007 {\it J. Phys. A: Math. Theor.} {\bf 40}
2935\hfill\break
\medskip
[3] Fu Q 1997 {\it Phys. Rev.}  A {\bf 56} 1806\hfill\break
\medskip
[4] Cohen-Tannoudji C, Dupont-Roc J and Grynberg G 1992 {\it
Atom-Photon Interactions: Basic Processes and Applications} (New
York: John Wiley \& Sons) \hfill\break
\medskip
[5] Adler S L and Bassi A, in preparation \hfill\break
\medskip
[6] Dalgarno  A and Kingston A E 1960 {\it Proc. Roy. Soc.} A {\bf
259} 424 \hfill\break
\medskip
[7] Panofsky W K H and Phillips M 1955 {\it Classical Electricity
and Magnetism} (Reading, MA: Addison-Wesley) \hfill\break
\medskip
[8] Pearle P 2007 private communication, based on Pearle P, Ring
J, Collar J I and Avignone F T 1999, {\it Found Phys} {\bf 29} 465
\hfill\break

\end{document}